\documentclass[%
 reprint,
superscriptaddress,
 amsmath,amssymb,
 aps,
floatfix,
]{revtex4-2}

\usepackage{dcolumn}
\usepackage{bm}
\usepackage{here}
\usepackage[pdftex]{graphicx}

\begin{document}


\title{$^1$H Polarization above 60\% at room temperature by triplet dynamic nuclear polarization}

\author{Kenichiro Tateishi}
\altaffiliation[kenichiro.tateishi@riken.jp]{}
\affiliation{Pioneering Research Institute, RIKEN, Saitama 351-0198, Japan}
\affiliation{Nishina Center for Accelerator-Based Science, RIKEN, Saitama 351-0198, Japan}

\author{Shuji Otsuka}
\affiliation{Pioneering Research Institute, RIKEN, Saitama 351-0198, Japan}
\affiliation{Department of Physics, Saitama University, Saitama 338-8570, Japan}

\author{Akihiro Yamaji} 
\affiliation{New Industry Creation Hatchery Center, Tohoku University, Miyagi 980-8579, Japan}

\author{Shunsuke Kurosawa} 
\affiliation{New Industry Creation Hatchery Center, Tohoku University, Miyagi 980-8579, Japan}
\affiliation{Institute of Laser Engineering, Osaka University, Osaka 565-0871, Japan}

\author{Tomohiro Uesaka} 
\affiliation{Pioneering Research Institute, RIKEN, Saitama 351-0198, Japan}
\affiliation{Nishina Center for Accelerator-Based Science, RIKEN, Saitama 351-0198, Japan}
\affiliation{Department of Physics, Saitama University, Saitama 338-8570, Japan}

\date{\today}

\begin{abstract}
  $^1$H polarization of 61\% was achieved by Dynamic Nuclear Polarization using photoexcited triplet electrons (Triplet-DNP) at room temperature and in 0.64 T. 
  We introduced dibenz[$\it {a,h}$]anthracene as a new host molecule of the polarizing agent, pentacene-$\it d$$_{14}$. 
  Its rigid structure provides a long spin-lattice relaxation time ($\it T$$_1$) of more than 2 hours at room temperature. 
  The single crystal of dibenz[$\it {a,h}$]anthracene doped with 0.05 mol\% pentacene-$\it d$$_{14}$ was grown by the Bridgman method, and cut into a small piece of $\sim$1 mg for Triplet-DNP experiment. 
  The $^1$H polarization buildup and relaxation measurements indicated that paramagnetic relaxation became the major source of the relaxation than spin-lattice relaxation.
  Finally, two promising applications of room-temperature hyperpolarization, $\it {i.e.}$ nuclear ordering and radiation-tolerant polarized target, are discussed.  
\end{abstract}

\maketitle

Nuclear spins have a great potential as the core of the next-generation quantum technologies.
Their long coherence time and good controllability are distinct features among a variety of quantum systems. 
Recently, they have been recognized as a new degree of freedom in solid-state physics, including magnetism \cite{Maekawa} and quantum information processing \cite{Urbaszek}. 
The effects on chemical and biological properties have been extensively investigated \cite{Vitalis, Fisher}. 
However, the small magnetic moment of the nuclei leads to significant disturbances from thermal energy, resulting in only slight $\it polarization$. 
Consequently, an ensemble of low purity of the states disappears its quantum properties.
The $^1$H polarization requires more than 40\% as well as a high density and strong interactions for the quantum applications~\cite{Abragam, Peres}.
Hyperpolarization of gaseous and solution materials have been achieved over 50\% using optical pumping~\cite{Chen} and the para-hydrogen induced polarization method~\cite{Korchak}, respectively.
This has never been accomplished in condensed materials at room temperature.

Dynamic nuclear polarization (DNP), a means of enhancing nuclear polarization by transferring that of electrons using microwave irradiation, is known to be the most promising for $^1$H polarization in solids~\cite{Abragam}.
It has been applied to polarized targets in particle~\cite{Andrieux} and nuclear physics~\cite{Uesaka}, too. 
Despite its capability to realize a close to unity ($\sim$100\%) $^1$H spin polarization, however, the DNP has a drawback that it demands extreme conditions; a high magnetic field ($>$2.5 T) and cryogenic temperatures ($<$1 K), to polarize the thermally-equilibrated electrons~\cite{Andrieux}.
In contrast, DNP using photoexcited triplet electrons (Triplet-DNP) can produce hyperpolarization under less demanding conditions \cite{Henstra,Takeda, Hamachi}. 
The triplet electrons have large non-equilibrated polarization, $\it {e.g.}$~$>$70\% for pentacene which is the standard polarizing agent of Triplet-DNP~(Fig.~1(a))~\cite{Ong}. 
The electron polarization originates from its molecular structure, independent of temperature and magnetic field, and enables nuclear poalrization at higher temperatures. 
The $^1$H spins in a single crystal of naphthalene doped with pentacene-$\it d$$_{14}$ were polarized up to 80\% at 25 K and in 0.3 T~\cite{Quan,Hautle}. 
$^1$H polarization of 34\% was obtained using a single crystal of $\it p$-terphenyl-$\it d$$_{4}$ doped with pentacene-$\it d$$_{14}$ at room temperature and in 0.4 T~\cite{Tateishi}.

Polarization at room temperature is bottlenecked by the relatively rapid spin relaxation of nuclei.
The relaxation is considered to be spin-lattice relaxation caused by molecular vibration and paramagnetic relaxation owing to triplet electrons~\cite{Tateishi}.
The breakthrough in the previous work was deuteration. 
The paramagnetic relaxation can be reduced by deuterating the polarizing agent~\cite{Eichhorn}. 
To suppress the spin-lattice relaxation, firstly, the mixture of 92.5\% $\it p$-terphenyl-$\it d$$_{14}$ and 7.5\% $\it p$-terphenyl was demonstrated, but the number of $^1$H spins decreased excessively~\cite{Kagawa}. 
The spin-lattice relaxation of the $\it p$-terphenyl crystal was mainly due to the pendulum motion of the central benzene ring, modulating the local dipolar field in and near the ring~\cite{Kohda}. 
Thus, a regioselective deuteration of the ring, $\it p$-terphenyl-$\it d$$_{4}$ with a deuteration ratio of 28.6\%, was developed~\cite{Tateishi}.
So far, the $^1$H polarization and the number of $^1$H spins were trade-off.

Here, we introduce the new host molecule, dibenz[$\it {a,h}$]anthracene~(Fig.~1).
The advantage of the molecule is to suppress the vibrational motion of the central benzene ring by fused-ring. 
For $\it p$-terphenyl crystal, the alignments of central rings for their associated outer rings are changed at 195 K~\cite{Kohda,David}. 
This phase transition has been characterized as an order-disorder transition. 
In the case of dibenz[$\it {a,h}$]anthracene, a similar transition occurred around 130 K~\cite{Zhao}, suggesting the vibrational mode is significantly suppressed. 
Surprisingly, the spin-lattice relaxation time ($\it T$$_1$) of the dibenz[$\it {a,h}$]anthracene crystal was longer than 2 hours at room temperature, and $^1$H polarization of 61\% was achieved at room temperature and in 0.64 T.
Finally, two promising applications for room-temperature hyperpolarization were discussed.
Note that another molecule, picene~\cite{Moro}, also has the rigid structure, but a highly-doped single crystal could not be grown because its melting point is higher than the decomposition temperature of pentacene.





\begin{figure}[t]
  \centering
  \includegraphics[width=1\columnwidth]{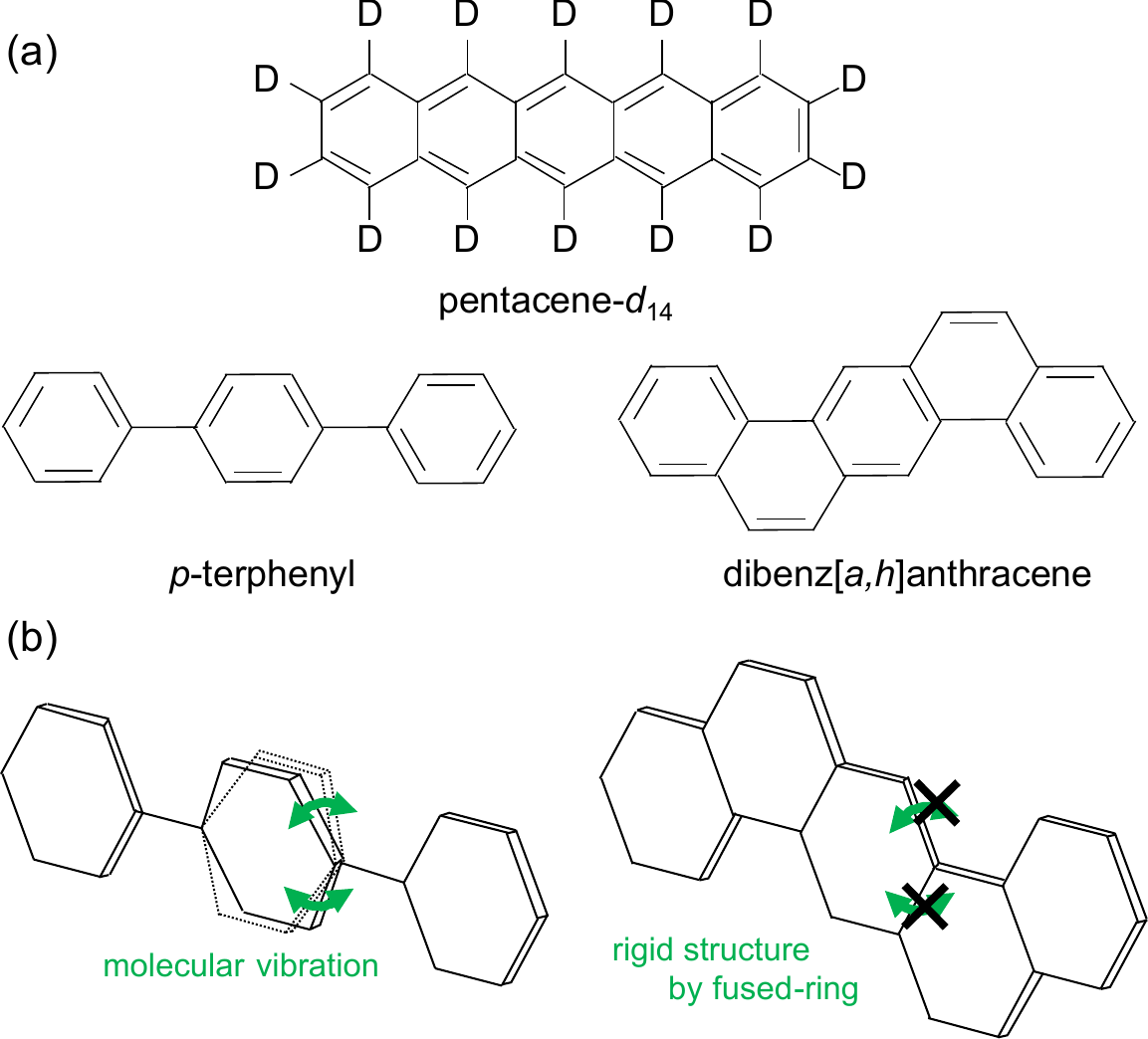}
  \caption{
  (a) Molecular structures of pentacene-$\it d$$_{14}$, $\it p$-terphenyl, and dibenz[$\it {a,h}$]anthracene. 
  (b) Schematics of $\it p$-terphenyl and dibenz[$\it {a,h}$]anthracene.
  The pendulum motion of central benzene ring in $\it p$-terphenyl shortens spin-lattice relazation time ($\it T$$_1$) by modulating the local dipolar field in and near the ring.
  That motion in dibenz[$\it {a,h}$]anthracene is surppressed by fused-ring.
  }
  \label{fig_1}
\end{figure}

Dibenz[$\it {a,h}$]anthracene and pentacene-$\it d$$_{14}$ were purchased from Sigma Aldrich and Toronto Research Chemical. 
Dibenz[$\it {a,h}$]anthracene was purified with the zone melting. 
Pentacene-$\it d$$_{14}$ was used without further purification. 
We fabricated the single crystal of dibenz[$\it {a,h}$]anthracene doped with 0.05 mol\% pentacene-$\it d$$_{14}$ by the Bridgman method (Fig. 2(a)).
The samples were sealed into a Pyrex ampoule. 
The temperature of the Bridgman furnace was set to 280~$^{\circ}$C, slightly higher than the melting point of dibenz[$\it {a,h}$]anthracene of 267~$^{\circ}$C. 
The growth rate of the crystal was set as $\sim$1.8~mm/h. 
A purple-red crystal was grown, but no birefringence was observed. 
Details of the crystal were analyzed using visible-light absorption (VIS), X-ray diffraction (XRD), and $^1$H-NMR.

\begin{figure}[b]
  \centering
  \includegraphics[width=1.0\columnwidth]{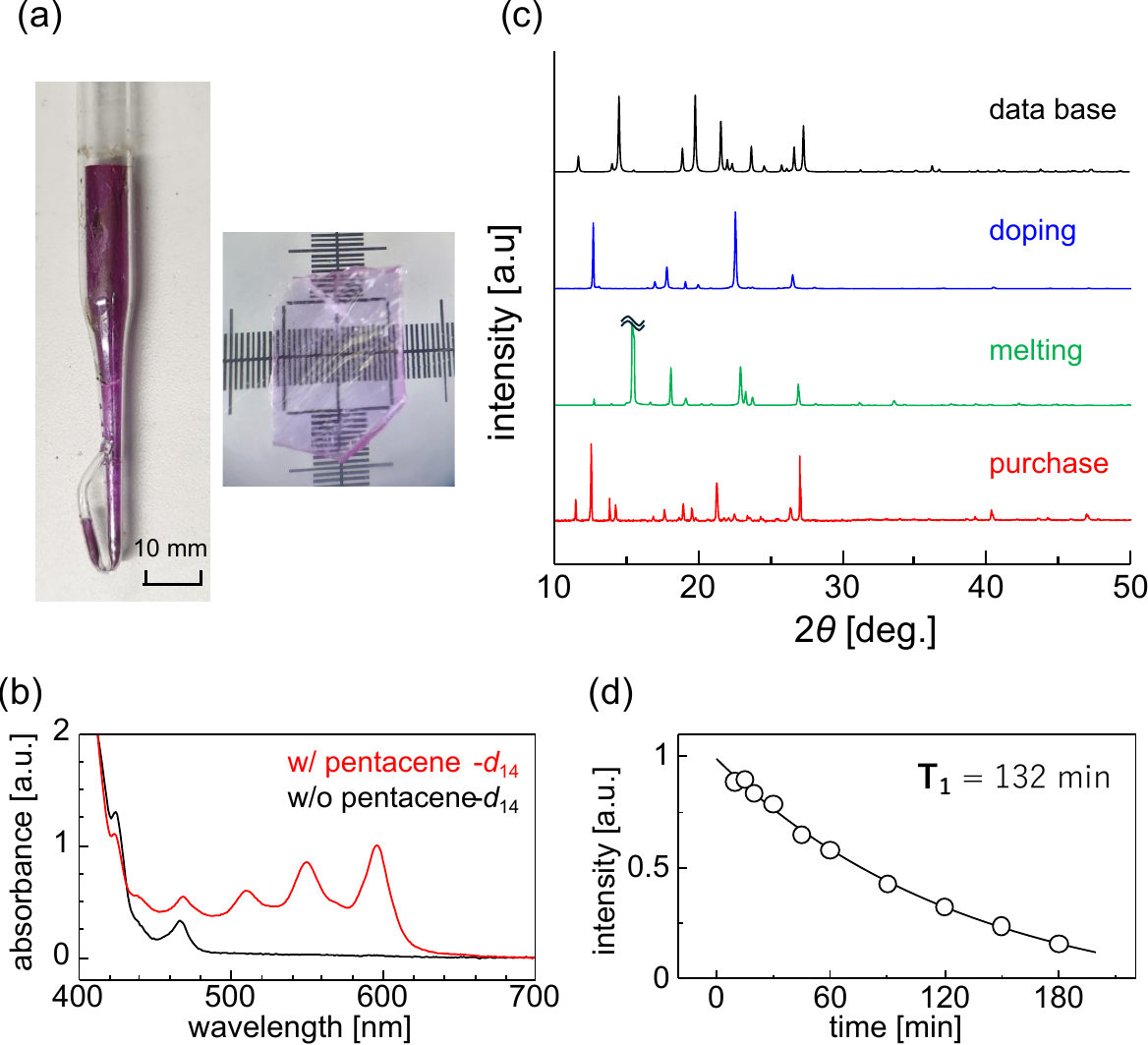}
  \caption{
    (a) Photos of Bridgman ampoule for production of a single crystal from its melt, and a single crystal of dibenz[$\it {a,h}$]anthracene doped with 0.05 mol\% pentacene-$\it d$$_{14}$. 
    (b) VIS spectra with/without 0.05 mol\% pentacene-$\it d$$_{14}$ doping. 
    (c) Power XRD patterns after grinding (red), zone melting (green), pentacene-$\it d$$_{14}$ doping (blue).
    They were much different from the data in Ref.~\cite{Iball} (black). 
    (d) $\it T$$_1$ of $^1$H spins in a single crystal of dibenz[$\it {a,h}$]anthracene doped with 0.05 mol\% pentacene-$\it d$$_{14}$.
    }
  \label{fig_2}
\end{figure}

The good dispersibility of pentacene-$\it d$$_{14}$ in the crystal was confirmed by the VIS spectra using the crystalline powders (Fig.~2(b)).
The planar structure of dibenz[$\it {a,h}$]anthracene allows pentacene doping comparable with $\it p$-terphenyl.
The absorption peaks were slightly red-shifted compared with the $\it p$-terphenyl crystal, showing the mono-dispersion of the pentacene-$\it d$$_{14}$ in dibenz[$\it {a,h}$]anthracene crystal. 
Remarkably, the dibenz[$\it {a,h}$]anthracene dominantly absorbs light below 480 nm.

The Cambridge Crystallographic Data Centre (CCDC) data base shows that dibenz[$\it {a,h}$]anthracene crystal (Database Identifier: DBNTHR02) has orthorhombic symmetry with four molecules per unit cell, the space group P$_{cab}$~\cite{Iball}. 
However, many anthracene derivatives have mechano-responsive properties~\cite{Gogoi, Kusukawa}.
In case of dibez[$\it {a,h}$]anthracene, the crystal structure was varied by grinding, zone melting, and even pentacene-$\it d$$_{14}$ doping.
The powder XRD patterns for pure and pentacene-$\it d$$_{14}$ doping samples were measured with a diffractometer (D8 DISCOVER, BRUKER) as shown in Fig. 2(c).
The results showed raw powder (as purchased) and melt-grown samples for the pure materials have different patterns, and the pentacene-$\it d$$_{14}$ doping sample after melting had different patterns.
Since its birefringence is hard to observe, the crystal axes were determined using the single-crystal XRD as follows; the crystal $\it a$-axis coincides with the laser beam direction and the $\it c$-axis is parallel to an external magnetic field.

The $^1$H spin relaxation was evaluated by measuring spin-lattice relaxation time ($\it T$$_1$) using $^1$H-NMR. 
The $\it T$$_1$ of a single crystal of dibenz[$\it {a,h}$]anthracene doped with 0.05 mol\% pentacene-$\it d$$_{14}$ was 132 min at room temperature and in 0.64 T (Fig. 2(d)).
To the best of our knowledge, this value is the longest as a pentacene-doped material at room temperature. 
We concluded that dibenz[$\it {a,h}$]anthracene, which has a planar and rigid molecule structure providing pentacene doping and a long $\it T$$_1$, is suitable for Triplet-DNP.

The procedure of Triplet-DNP was as follows.
Pentacene-$\it d$$_{14}$ was photoexcited with a 545 nm pulsed laser with a width of 1 $\mu$s.
Then, a 17.2 GHz microwave with a width of 20 $\mu$s was irradiated to transfer the polarization from electron spins to $^1$H spins by matching the Rabi frequency of the electron spin with the Larmor frequency of the $^1$H spins. 
The magnetic field was swept during microwave irradiation to improve the transfer efficiency. 
This scheme was called the Integrated Solid Effect (ISE) \cite{Henstra14}. 
By repeating the ISE sequence, the $^1$H polarization was buildup until it balanced with the spin relaxation.
The time evolution of $^1$H polarization $\it P$$_{(t)}$ with Triplet-DNP is phenomenologically represented by the following equation:
\begin{gather}
  \frac{dP_{(t)}}{dt} = \frac{1}{T_D}(P_e-P_{(t)})-\frac{1}{T_R}(P_{(t)}-P_{th}), \\
  \frac{1}{T_R} = \frac{1}{T_1} + \frac{1}{T_e},
  \label{rate}
\end{gather}
where $\it T$$_D$ and $\it T$$_R$ are the buildup and the spin relaxation time constants, respectively. 
The $\it P$$_e$ and $\it P$$_{th}$ represent the electron and the thermally-equilibrated nuclear polarization, respectively. 
The $\it T$$_R$ can be decomposed into two effects of spin-lattice relaxation time ($\it T$$_1$) caused by molecular vibration and paramagnetic relaxation time ($\it T$$_e$) owing to triplet electrons \cite{Tateishi}. 
Since the $\it P$$_{th}$ is small as $\sim$0.0001\% in our experimental conditions, Eq. (1) is solved by omitting it: 
\begin{equation}
  P_{(t)} = \frac{P_e}{1+\frac{T_D}{T_R}}\bigg\{1-{\rm exp}\bigg[-t\bigg(\frac{1}{T_D}+\frac{1}{T_R}\bigg)\bigg]\bigg\}.
  \label{buildup_eq}
\end{equation}
The final attainable polarization is given as $P_e/(1 +T_D/T_R )$, indicating the suppression of the relaxation is critical for higher polarization.

Triplet-DNP was carried out at room temperature and in 0.64 T. 
A single crystal of dibenz[$\it {a,h}$]anthracene doped with 0.05 mol\% pentacene-$\it d$$_{14}$ with the weight of $\sim$1 mg was utilized.
Details of the experimental setup were described in Ref.~\cite{Kouno}. 
After repeating the ISE sequence with a frequency of 1 kHz for a certain time, the sample was shuttled into an NMR coil within 1 s, and $^1$H NMR spectra were measured. 
The $^1$H polarization was calculated by comparing a reference signal of ethanol. 
Figure 3 shows the $^1$H polarization buildup curve as a function of time. $^1$H polarization after 2.5 hours triplet-DNP reached 61\% with the buildup constant of 20.2 min. 
This is the highest ever $^1$H polarization achieved in solids at room temperature. 
We also measured the $\it T$$_1$ under laser irradiation ($\it T$$_R$) by changing the time from Triplet-DNP to NMR detection. 
$\it T$$_R$ with a laser frequency of 1 kHz was 57.1 min. 
Using Eq. (2), $\it T$$_e$ were calculated to be 96.9 min. 
Comapring with the $\it T$$_1$ of 132 min in Fig. 2(d), paramagnetic relaxation became a major contribution of the relaxation, indicating sample cooling to prolong $\it T$$_1$ by suppressing molecular vibrations might no longer be effective. 
The other polarizing agents with a shorter triplet lifetime, diaza-substituted pentacene and tetracene, have been developed \cite{Kouno}. 
The molecules with deuteration are expected to have a longer $\it T$$_e$, consequently, the final attainable $^1$H polarization would be increased.
  
\begin{figure}[t]
  \centering
  \includegraphics[width=0.8\columnwidth]{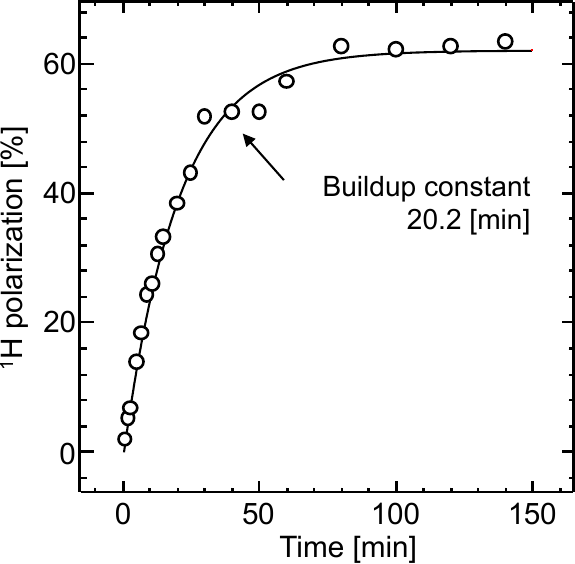}
  \caption{
    $^1$H polarization buildup curve of the single crystal of dibenz[$\it {a,h}$]anthracene doped with 0.05 mol\% pentacene-$\it d$$_{14}$ at room temperature and 0.64 T. Tripelt-DNP was applied with a frequency of 1 kHz. 
  }
  \label{fig_3}
\end{figure}

The hyperpolarization of $>$60\% achieved at room temperature can open two promising applications.

1) Nuclear ordering at room temperature:
The electronic magnetism has been a main topic of solid-state physics. 
Compared to that, strongly correlated nuclear spin systems have been rarely studied due to their weak interaction. 
Appearance of quantum properties such as magnetic phase transition or entanglement, requires the the $^1$H polarization higher than 40\%~\cite{Abragam, Peres}.
Interestingly, nuclear ordering induced by the dipolar interaction can be different from electronic ordering owing to the exchange interaction, since the dipolar interaction is long-ranged and the sign can vary the relative position between two spins, $\it {i.e.}$ sample rotation.
The correlated spins can also affect non-magnetic properties, such as photoluminescence~\cite{Kandrashkin} and thermal conductivity~\cite{Kawamata}. 
At room temperature, it enables the combination of nuclear magnetism with those measurements, expecting the creation of new quantum materials utilizing nuclear polarization as a degree of freedom.

2) Radiation-tolerant polarized target:
Hyperpolarization is also applied as a polarized target in accelerator sciences~\cite{Andrieux,Uesaka}. 
This provides us with unique information as spin-dependent observables. 
However, high-energy ion beams often depolarize the target by accelerating the relaxation, called radiation damage. 
It mainly originates from the chemical bond breaking of the target material, generating unwanted radicals. 
This can be significantly reduced at high temperature, known as annealing effect~\cite{McKee}. 
In the case of organic crystals, the annealing occurs at temperature higher than 190 K~\cite{Capozzi} and the damage is expected to recover spontaneously. 
Thus, room-temperature hyperpolarization benefits not only the simplification of experimental setup but also the limitation of beam intensity, $\it {i.e.}$ statistical accuracy of the experiment.

In summary, $^1$H polarization of 61\% was achieved by Triplet-DNP at room temperature and in 0.64 T. 
For the room-temperature hyperpolarization, the bottleneck was the spin- relaxation of nuclei. 
We introduced a new host molecule, dibenz[$\it {a,h}$]anthracene, of the pentacene-$\it d$$_{14}$ to suppress the vibrational motion, instead of deuteration. 
The single crystal of dibenz[$\it {a,h}$]anthracene doped with 0.05 mol\% pentacene-$\it d$$_{14}$ was grown by the Bridgman method. 
The planar structure of dibenz[$\it {a,h}$]anthracene allowed pentacene doping, and VIS and Power XRD spectra agreed with a mono-dispersion in the crystal. 
In addition, the rigid structure provided a long $\it T$$_1$ of 132 min at room temperature. 
The $^1$H polarization buildup and relaxation measurements indicated that paramagnetic relaxation became the major source of the relaxation than spin-lattice relaxation.
Room-temperature hyperpolarization has the potential to apply nuclear ordering and radiation-tolerant polarized target.
Moreover, the hyperpolarization of nuclear spins is expected to open the door to next-generation quantum technologies.

This work was partly supported by the RIKEN Cluster for Science, Technology and Innovation Hub (RCSTI), the RIKEN TRIP initiative (Fundamental Quantum Science Program), GIMRT Program of the Institute for Materials Research, Tohoku University (Proposal No. 202208-RDKYA-0090, and 202305-RDKGE-0095), JSPS KAKENHI (JP24K08471, and JP20H05636), and JST ERATO Grant No. JPMJER2304, Japan.

\end{document}